\documentclass[twocolumn,aps,superscriptaddress,showpacs,preprintnumbers,amsmath,amssymb]{revtex4-1}
\usepackage{graphicx} 
\usepackage{bm} 

\newcommand{\beq}{\begin{equation}}
\newcommand{\eeq}{\end{equation}}

\def\vec#1{\mathbf{#1}}

\def\xi{\mathbf{x}_i}

\begin{document}

\title{Tunable exciton interactions in optical lattices with polar molecules}

\author{Ping Xiang, Marina Litinskaya, and Roman V. Krems}

\address{Department of Chemistry, University of British Columbia, Vancouver, V6T 1Z1, Canada}

\begin{abstract}
Rotational excitation of polar molecules trapped in an optical lattice gives rise to rotational excitons. Here we show that non-linear interactions of such excitons can be controlled by an electric field. The exciton--exciton interactions can be tuned to induce exciton pairing, leading to the formation of biexcitons. Tunable non-linear interactions between excitons can be used for many applications ranging from the controlled preparation of entangled quasiparticles to the study of polaron interactions and the effects of non-linear interactions on quantum energy transport in molecular aggregates. 
\end{abstract}

\maketitle

{\it Introduction.}  
The absorption of photons by a solid-state crystal gives rise to quasiparticles called excitons. There are two limiting models of excitons: Wannier-Mott excitons and Frenkel excitons. Wannier-Mott excitons occur in crystals with band structure leading to collective excitations with an effective radius much greater than the lattice constant, while Frenkel excitons are typical for molecular crystals, where collective excitations are superpositions of elementary excitations localized on different lattice sites. These properties lead to important differences in non-linear exciton interactions for the two models. The interactions between the Wannier-Mott excitons are determined by the Coulomb interaction and the phase space filling \cite{nolinear-wannier}, while the interactions of Frenkel excitons are determined by shorter range dynamical couplings \cite{agranovich}. Multiple experiments demonstrated that Wannier-Mott excitons can form two-exciton bound states called biexcitons \cite{wmxx,stevenson2006, Lozovik2002}. By contrast, despite many theoretical studies  \cite{vektaris, Ezaki1994, biexciton-theory-1,biexciton-theory-2}, Frenkel biexcitons have eluded the experimental observation, with one notable exception \cite{frenkelxx}. In the present work, we show that rotational excitation of ultracold molecules trapped on an optical lattice gives rise to Frenkel excitons with controllable non-linear interactions. We demonstrate that the exciton--exciton interactions can be tuned to induce the formation of Frenkel biexciton and that the biexciton binding energy can be controlled by an external electric field.    

Several experiments have recently demonstrated that ultracold molecules can be trapped in a periodic potential of an optical lattice \cite{umol}. Such systems can be used for the study of quantum energy transfer \cite{felipe, scholes2006},  non-linear photon--photon interactions \cite{suzanne}, novel quantum memory devices \cite{peter-rabl} and, most notably, for quantum simulation of lattice models \cite{Barnett2006, micheli2006, Carr, Trefzger2010, Kestner2011, gorshkov}. Although describing different phenomena, the Hamiltonians presented in these references, and in the present work, can be cast in the same form. For example, the exciton Hamiltonian discussed here can be mapped onto the $t$-$V$ model, which is the special case of the Heisenberg-like models studied in the context of ultracold molecules in Refs. \cite{micheli2006,gorshkov}. The key difference of the present work from those in Refs. \cite{micheli2006,gorshkov} is that we explore phenomena associated with the excitation spectrum of the many-body system in the limit of a small number of excitations. Because we consider a simpler Hamiltonian, our scheme is conceptually simpler, requiring fewer molecular states and external field parameters.  We use the rotational states of molecules as a probe of the collective interactions, i.e. the experiments proposed here can be carried out by measuring site-selective populations of the rotational states. This can be achieved by applying a gradient of an electric field and detecting resonant transitions from Stark-shifted levels, as described in Ref. \cite{demille}.

{\it Exciton--exciton interactions in an optical lattice.}
We consider an ensemble of polar diatomic molecules in the $^1\Sigma$ electronic state trapped on an optical lattice 
in the ro-vibrational ground state. The rotational states of the molecules $|NM_N\rangle$ are described by the rotational angular momentum 
$\hat{N}$ and its projection on the quantization axis $M_{N}$. We assume that the molecules are in the Mott-insulator phase \cite{umol} and that each lattice site contains only one molecule.  
We consider the rotational excitation $|N=0, M_N = 0 \rangle \rightarrow |N = 1, M_N = 0 \rangle$ of molecules in the lattice \cite{note}.
For simplicity, we denote the ground state of the molecule in site $n$ by $|g_{n}\rangle$ and the excited state by $|e_{n}\rangle$. 
 Because the molecules are 
coupled by the dipole-dipole interaction, the rotational excitation gives rise to a rotational Frenkel exciton \cite{felipe}, which is an eigenstate of the Hamiltonian: 
\begin{equation}
\hat{H}_{\rm exc} = E_0 \sum\limits_{n=1}^{N_{\rm mol}} \hat{P}_n^\dag
\hat{P}_n +  \sum\limits_{n,m \neq n}^{N_{\rm mol}}
J(n-m)\hat{P}_n^\dag \hat{P}_m,
\label{exciton}
\end{equation}
where $J(n-m) = \langle e_{n} , g_{m} | \hat{V}_{dd}(n-m) | g_{n}, e_{m} \rangle$ with $\hat{V}_{dd}(n-m)$ representing dipole-dipole interaction
between molecules in sites $n$ and $m$, $E_0$ is the energy difference between the states $|g\rangle$ and $|e\rangle$, 
and the operators $\hat{P}_n^\dag$ and $\hat{P}_n$ are defined by the relations 
$\hat{P}_n^\dag | g_{m} \rangle = \delta_{nm} | e_{n}
\rangle$ and $\hat{P}_n | e_{m} \rangle = \delta_{nm} | g_{n} \rangle$.
The $\hat{P}_n$ and $\hat{P}_m$ operators satisfy the bosonic commutation, if $n \neq m$, and the fermionic commutation, if $n = m$ \cite{agranovich}.

Multiple excitations lead to dynamical exciton--exciton interactions described by \cite{agranovich}:
\begin{equation}\label{dynamical}
\hat{H}_{\rm dyn} = \frac{1}{2} \sum\limits_{n,m \neq n}^{N_{\rm mol}} D(n-m)
\hat{P}_{n}^\dag \hat{P}_{m}^\dag \hat{P}_{n} \hat{P}_{m} 
\end{equation}
where $D(n - m) = \langle e_{n}, e_{m} | \hat{V}_{dd}(n-m) | e_{n}, e_{m} \rangle +
\langle g_{n}, g_{m} | \hat{V}_{dd}(n-m) | g_{n}, g_{m} \rangle - 2\langle e_{n}, g_{m} |
\hat{V}_{dd}(n-m) | e_{n}, g_{m} \rangle$. The dipole - dipole interaction operator $\hat{V}_{dd}(n-m)$ can only couple states of different parity \cite{RotSpect}. If $|g \rangle$ and $| e \rangle$ are states of well-defined parity, such as the rotational states $|NM_N \rangle$, the matrix elements $D(n-m)$ must be zero. 

 The inversion symmetry (parity) of molecules on an optical lattice can be broken by applying an external dc electric field. In an electric field, $|g_{n} \rangle$ and $| e_{n} \rangle$ are eigenstates of the Hamiltonian 
$\hat{H}^{\rm mol}_n = B_e \hat{N}_n^2 - {\vec d}_n \cdot  {\vec {\cal E}_f}$, where ${\vec d}_n$ is the dipole moment of molecule in site $n$ and ${\vec {\cal E}_f}$
is the electric field vector. They can be expressed as $| g \rangle = \sum_N a_N |N M_N = 0 \rangle$ and $| e \rangle = \sum_N b_N |N M_N = 0 \rangle$, where 
 $a_N$ and $b_N$ are determined by the electric field strength.  
 
 The Hamiltonian (\ref{exciton}) can be diagonalized by the Fourier transforms: $\hat{P}^\dag(k) = \frac{1}{\sqrt{N_{\rm mol}}}
\sum\limits_n e^{ikn} \hat{P}_n^\dag$ and $\hat{P}(k) =
\frac{1}{\sqrt{N_{\rm mol}}} \sum\limits_n e^{-ikn} \hat{P}_n$,  where
$k$ is the wave vector of the exciton.  This transformation leads to 
 $\hat{H}_{\rm exc} = \sum\limits_k E(k)
\hat{P}^\dag(k) \hat{P}(k)$, where $E(k) = E_0 + J(k)$ with $J(k) =
\sum\limits_n J(n) e^{-ikn}$, and  $\hat{P}^\dag(k)$ and $\hat{P}(k)$ 
create and annihilate Frenkel excitons with energies $E(k)$.  The interaction (\ref{dynamical})
in the momentum representation is: 
\begin{equation}\label{dynamical-k}
\hat{H}_{\rm dyn} = \frac{1}{N_{\rm mol}} \sum\limits_{k_1, k_2, q} D(q)
\hat{P}^\dag(k_1+q) \hat{P}^\dag(k_2-q) \hat{P}(k_1) \hat{P}(k_2),
\end{equation}
where $D(q) = \sum_n D(n) e^{-iqn}$.

{\it Biexcitons.}
The exciton--exciton interactions generally have little effect on the energy spectrum of two-particle
continuum states $E(k_1) + E(k_2)$. However, under certain conditions discussed below, non-linear
interactions may result in the formation of a bound two-exciton
complex, biexciton. The biexciton state is split from the
two-particle continuum. The splitting is the biexciton binding energy.

Ref. \cite{vektaris} shows that biexcitons can generally form in 1D and 2D crystals if
\begin{equation}\label{biexc-formation}
|D| \geq 2 |J|
\end{equation}
where $D$ and $J$ are the coupling constants $D(n-m)$ and $J(n-m)$ for $n-m =1$. In 3D crystals, biexcitons can form if $|D| > 6 |J|$ \cite{biexciton-theory-2}. For 1D crystals, the biexciton energy in the nearest neighbor approximation (NNA) is
$E_b(K) = 2E_0 + D + \frac{4 J^2 \cos^2(aK/2)}{D}$,
where $K$ is the total wave vector of two interacting excitons and $a$ is the lattice constant, and the biexciton binding energy is $\Delta = (D - 2J)^2/D$. The maximum number of exciton--exciton bound states is equal to the dimensionality of the crystal, i.e. one biexciton state for 1D, two for 2D and three for 3D \cite{vektaris,note1}. 

\begin{figure}[ht]\label{f-D/J}
\centering
\includegraphics[width=\linewidth]{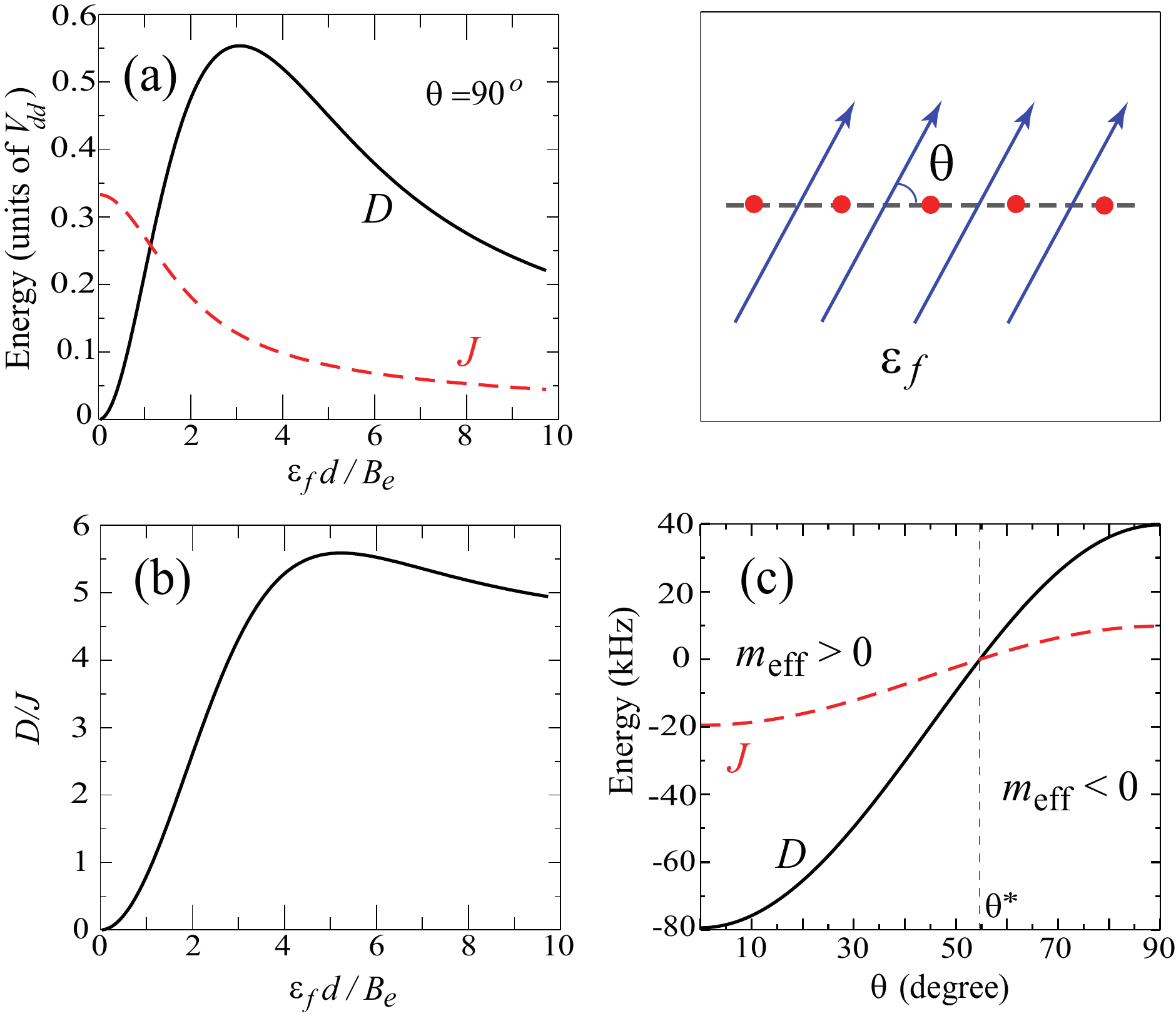}
\caption{(Color online) (a) The parameters $D$ and $J$ (in units of $V_{dd} = d^2/a^3$) as functions of the electric field strength. (b) The ratio  $D/J$ as a function of the electric field strength. The field is perpendicular to the intermolecular axis.
For LiCs molecules possessing the dipole moment $d$=5.529~Debye, the value ${\cal E}_f d/B_e = 1$ corresponds to ${\cal E}_f = 2.12$ kV/cm. (c) Schematic depiction of the angle $\theta$ between the field (represented by blue arrows) and the molecular array (represented by red dots). (d) $D$ and $J$ for a 1D array of LiCs molecules separated by 400 nm as functions of $\theta$ for ${\cal E}_f = 6$ kV/cm.
}
\end{figure}

\begin{figure}[ht]\label{f-D/J-spectra}
\centering
\includegraphics[width=\linewidth]{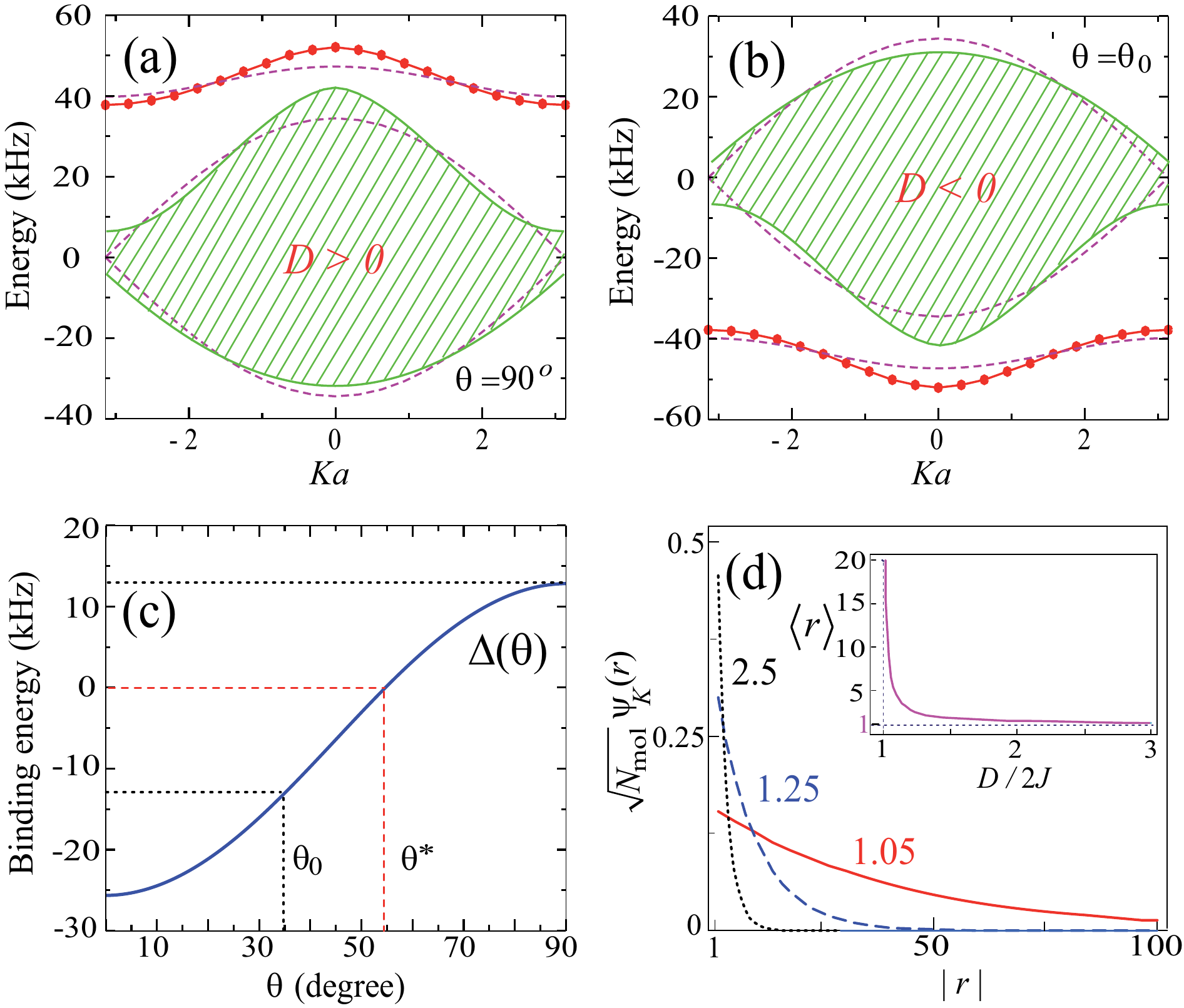}
\caption{(Color online) (a) and (b): Two-excitation spectra of a 1D array of LiCs molecules on an optical lattice: NNA
(dashed lines) and exact solutions (solid lines). The shaded regions encapsulate the bands of the continuum two-exciton states. (c)
$\theta$-dependence of the biexciton binding energy $\Delta$. The electric field magnitude is 6.88 kV/cm, $\theta_0 = \arccos \sqrt{2/3}$, $\theta^\ast = \arccos \sqrt{1/3}$. (d) Biexciton wave function vs the lattice site separation $|r|=|n-m|$ of the two excitations for $K=0$. Inset: Mean width of the biexciton wave function $\langle r \rangle$ calculated as the width of $\psi^{2}_{K}(r)$ at half maximum. Numbers at each curve indicate the value of $D/2J$.}
\end{figure}

For molecules in an optical lattice, the magnitudes of $J$ and $D$ depend on the strength of the applied electric field and the angle between the field and the molecular array  ($\theta$) (see Figure 1c).  We calculate these parameters for a 1D array of $^1\Sigma$ polar molecules (such as alkali metal dimers produced in ultracold molecule experiments) trapped on an optical lattice with the lattice separation $a = 400$ nm. Figure 1b shows that for a fixed angle $\theta$ the ratio $D/J$ increases as the electric field magnitude increases. 
For LiCs molecules, the condition (\ref{biexc-formation}) is satisfied for electric fields $>3.6$~kV/cm. Note that the ratio $D/J$ is independent of $\theta$. 

 Frenkel excitons are quasiparticles characterized by an effective mass ($m_{\rm eff}$). The sign of $J$ determines the sign of the  effective mass \cite{felipe}: negative $J$ corresponds to positive $m_{\rm eff}$ and vice versa (see Figure 1d). Due to the linearity of the Schr\"{o}dinger equation, a positive potential is attractive for particles with negative mass, just like a negative potential is attractive for particles with positive mass.  Because the sign of $J$ and $D$ is the same (and consequently the signs of $D$ and $m_{\rm eff}$ are opposite), the dynamical interaction (\ref{dynamical}) between excitons in this system is attractive.

To demonstrate the formation of the biexciton and calculate the biexciton energy, we diagonalize the Hamiltonian $\hat{H}_{\rm exc} + \hat{H}_{\rm dyn}$ for a one-dimensional array of $N_{\rm mol} = 501$ LiCs molecules. The matrix of the Hamiltonian is evaluated by expanding the biexciton states as 
\begin{eqnarray}
| \Psi_b(K) \rangle = 
\sum\limits_{k \geq 0} C_k^K \ 
| \hat{P}^\dag(K/2 + k)  \hat{P}^\dag(K/2
- k) \rangle,
\label{wave-function} 
\end{eqnarray}
where $K = k_1+k_2$ and $k = (k_1 - k_2)/2$, and $k_1$ and $k_2$ denote the wavevectors of the interacting excitons. The Hamiltonian matrix is diagonalized numerically for fixed values of $K$, which is conserved. Figure 2 shows that for $\theta = 90^o > \theta^*=\arccos(1/\sqrt{3})$ the biexciton energy is above the two-exciton continuum (binding for particles with negative mass), and for $\theta = \arccos \sqrt{2/3} <  \theta^*$ below it (binding for particles with positive mass). The binding energy of the biexciton changes sign at $\theta = \theta^*$. 
It is instructive to analyze the biexciton wave function in the site representation, which can be obtained analytically using NNA:
\begin{eqnarray}
&&| \Psi_b(K) \rangle = \sum\limits_{n,m \neq n} e^{iK(n+m)/2}
\psi(|n-m|) | \hat P_n^\dag \hat P_m^\dag \rangle , \nonumber \\
&&\psi_{K}(r) = \frac{\sqrt{D^2 - 4 J^2\cos^2 (Ka/2)}}{2D \sqrt{N_{\rm  mol}}} \left(
\frac{2 J\cos(Ka/2)}{D} \right)^{|r|-1} , \nonumber
\end{eqnarray}
where $r=n-m$.
The biexciton wave function $\psi_{K}(r)$ is plotted in Figure 2d.
Figures 1 and 2 thus illustrate that the biexciton binding energy and size can be tuned by varying the strength and orientation of the electric field.

{\it Non-optical creation of biexcitons.} 
We now explain why it is difficult to observe Frenkel biexcitons in solid-state molecular crystals. First, many molecular crystals, such as anthracene or naphthalene,  possess inversion symmetry. In these crystals, the constant $D$ as defined in the line after Eq.(\ref{dynamical}) must vanish and Eq.(\ref{biexc-formation}) is not satisfied. Second, it is difficult to excite biexciton states optically: it was shown in Ref.\cite{Ezaki1994} that the oscillator strength for the photon-induced transitions to the biexciton state must decrease with the increasing binding energy of the biexcitons. Therefore, two-photon excitation can only produce unstable weakly bound biexcitons. Third, excitons in molecular crystals decay via bimolecular annihilation processes into higher-energy states and subsequent relaxation accompanied by emission of phonons. This process is prohibited by conservation of energy in an optical lattice with diatomic molecules. Figure 1 demonstrates that the first obstacle can be removed by tuning the electric field. To overcome the second obstacle, we propose a non-optical method of populating deeply bound biexciton states based on the unique structure of $^1\Sigma$ polar molecules. 
As zero electric field the rotational states $|g\rangle$ and $|e\rangle$ are separated by the energy $\Delta \epsilon_{e - g}= 2 B_e$, while the energy separation between state $|e\rangle$ and the next rotationally excited state $|f\rangle \equiv |N=2, M_N = 0\rangle$ is equal to $\Delta \epsilon_{e- f}=4B_e$. As the electric field increases, $\Delta \epsilon_{e - g}$ increases faster than $\Delta \epsilon_{e - f}$.
When ${\cal E}_f d/B_e \simeq 3.24$ (corresponding to ${\cal E}_f \simeq 6.88$ kV/cm for LiCs), $\Delta \epsilon_{f - g} = 2 \Delta \epsilon_{e - g}$. At electric fields near this magnitude, two $| g \rangle \rightarrow | e \rangle$ excitons can undergo the transition to the $|f \rangle$ state, and, inversely, the $|g \rangle \rightarrow |f\rangle$ excitation can produce a pair of $|g\rangle \rightarrow |e \rangle$ excitons or a biexciton state depicted in Figure 2. The coupling between states $| e \rangle$ and $| f \rangle$ is
$\hat{H}_{12} = \sum\limits_{n \neq m} M(n-m) \hat{R}_n  \hat{P}_n^\dag
\hat{P}_m^\dag$, where $M(n-m) = \langle e_{n},e_{m} | V_{dd}(n-m) | f_{n},g_{m}
\rangle$, and the operator $\hat{R}_n$ annihilates the $| f \rangle$
excitation in lattice site $n$. The total Hamiltonian describing this three-level 
system is $\hat{H}_{g-e-f} = \hat{H}_{\rm exc} + \hat{H}_{\rm dyn} +  \hat{H}_2 + \hat{H}_{12}$, where
$\hat{H}_2 = E_f \sum\limits_n \hat{R}_n^\dag \hat{R}_n   + \sum\limits_{n,m \neq n} J_{g-f}(n-m) \hat{R}_n^\dag \hat{R}_m $ and $J_{g-f}(n-m) = \langle g_{n},f_{m} | V_{dd}(n-m) | f_{n},g_{m} \rangle$.

\begin{figure}[ht]\label{f-population dynamics}
\centering
\includegraphics[scale=0.5]{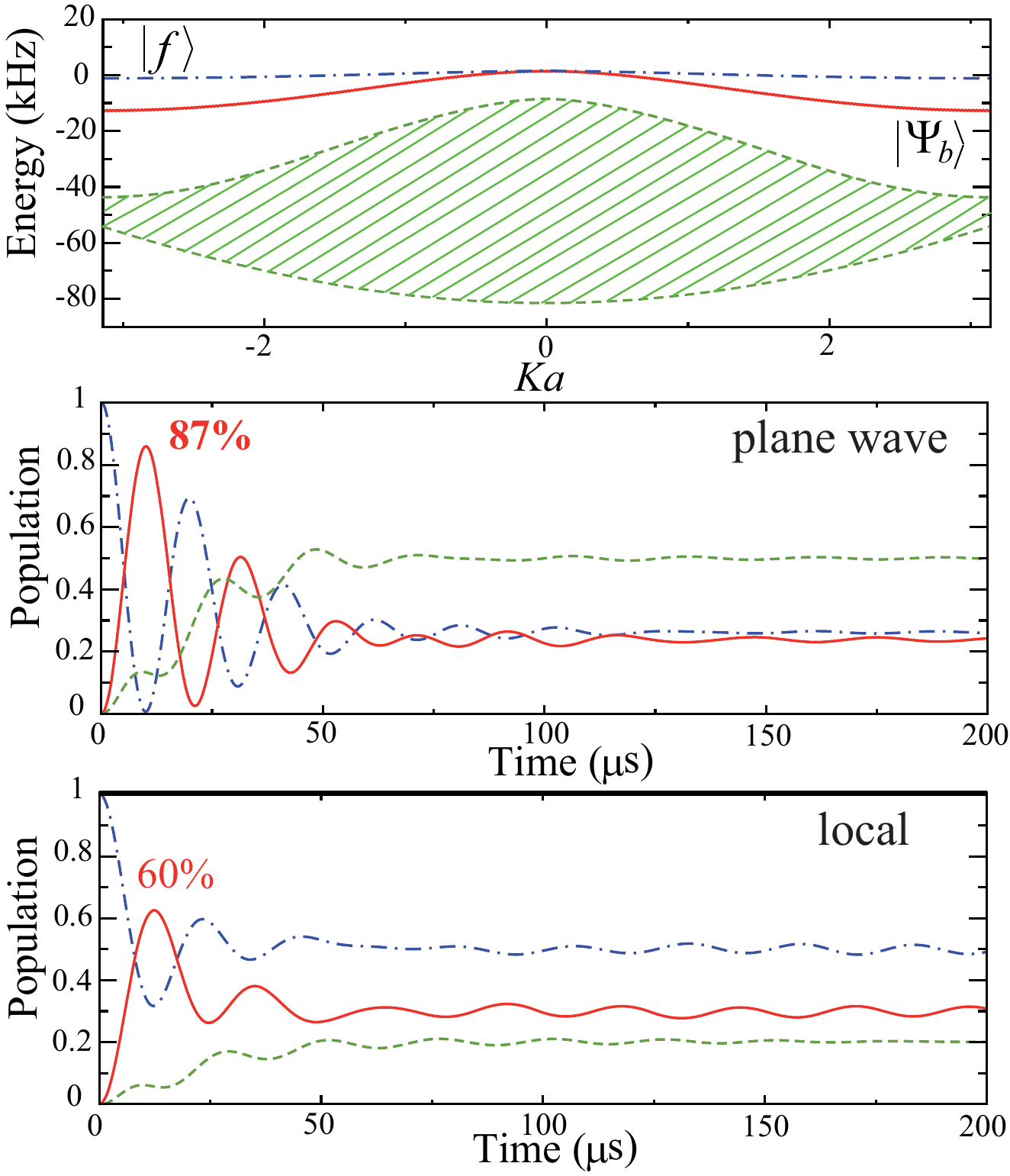}
\caption{(Color online) Population dynamics for the transition from $|g\rangle \to |f\rangle$ exciton (middle panel) and from an $f$ state localized on a single molecule (lower panel) to coherent $|g\rangle \to |e\rangle$ excitons and biexcitons. The green dashed curves denote the population accumulated in the pairs of non-bound $|g\rangle \to |e\rangle$ exciton states, the red solid curves the biexciton state and the blue dot-dashed curves the $f$ state. The shaded region in the upper panel encapsulates the band of the continuum two-exciton states. The calculation is for a 1D ensemble of $N_{mol}=501$ LiCs molecules on an optical lattice with lattice separation a = 400 nm. The electric field of magnitude 6.88 kV/cm is perpendicular to the molecular array.}
\end{figure}

In order to calculate the probability of the population transfer from state $f$ to the biexciton state, we solve the time-dependent Schr\"{o}dinger equation with the Hamiltonian $\hat{H}_{g-e-f}$ evaluated in the basis of products of the eigenstates of $\hat{H}_{\rm exc} + \hat{H}_{\rm dyn}$ and the eigenstates of $\hat{H}_2$. This leads to coupled differential equations, which we solve numerically. Note that the magnitude of $J_{g-f}$ is about ten times smaller than $J$.
 In the absence of decoherence, the $| g \rangle \rightarrow | f \rangle$ excitation gives rise to the Frenkel exciton and the transition from the $f$ states to 
 the biexciton state is a coherent exciton--exciton transition. In the presence of decoherence, the exciton states become localized. If the decoherence rate is larger than $J_{g-f}/h$, but smaller than $J/h$, the $| g \rangle \rightarrow | f \rangle$ excitation is localized, while the biexciton states remain coherent. Figure 3 presents the calculations of the population transfer probabilities for both scenarios. 
The results show that the biexciton states can be populated with high efficiency. The equilibrium populations (in the limit of large $t$)
depend on the relative energies of the $f$ state, the biexciton bound state and exciton--exciton continuum states, which can be tuned by varying the electric field magnitude. 
The efficiency of the population transfer can be maximized if the electric field is detuned far away from resonance when the biexciton population oscillations reach the first maximum. Detuning the electric field to low magnitudes effectively decouples the $f$ state from the states in the $\{g,e\}$ subspace and interrupts the population dynamics. This corresponds to switching off the channel for bimolecular annihilation of excitons, which is one of the reasons of the biexciton population depletion in solids. We have confirmed that the calculations with electric fields $< 5.0$ kV/cm yield no  noticeable population transfer.

{\it Discussion.} We have shown that rotational excitation of molecules trapped on an optical lattice gives rise to rotational excitons whose interactions can be controlled by an external electric field. The exciton--exciton interactions can be tuned to produce two-exciton bound states.  A biexciton is an entangled state of two Frenkel excitons. The creation of biexcitons as described in the previous section and tuning the electric field to the regime of zero binding energy can thus be used for the controlled preparation of entangled pairs of non-interacting excitons. In order to observe the biexcitons, one could measure correlations between the populations of the rotationally excited states of molecules in different lattice sites using the method proposed in Ref. \cite{demille}.

The present work suggests several interesting questions. For example, it was recently shown that Frenkel excitons in shallow optical lattices can be coupled to lattice phonons, leading to polarons \cite{felipe-polarons}. Coupling a Frenkel biexciton to phonons would produce strongly interacting polarons. It would be interesting to explore if these interactions lead to the formation of bipolarons.

We have repeated the calculations presented here for a system of three excitons and similarly observed the formation of three-exciton bound states. It would be interesting to explore the effect of tunable exciton--exciton interactions on excitation correlations, both as a function of $D/J$ and the density of excitations, to understand fundamental limits of exciton clustering \cite{exciton-cluster}.

The creation of biexcitons with tunable binding energy and measuring quantum energy transport for different ratios $D/J$ can be used to study the effects of exciton--exciton entanglement on energy transfer in molecular aggregates \cite{photosynthesis-1,photosynthesis-2,photosynthesis-3}. The ability to tune exciton--exciton interactions can be used to explore the role of multiple excitation correlations on energy transfer in disordered systems (the confining lattice potential can be tilted or the molecules can be perturbed by a disorder potential produced by an inhomogeneous electric field).
  
{\it Acknowledgment.} We thank Felipe Herrera for useful discussions. RVK acknowledges support from the Institute for Theoretical, Atomic, Molecular and Optical Physics at the Harvard-Smithsonian Center for Astrophysics in the form of a sabbatical fellowship. This work was supported by NSERC of Canada and the Peter Wall Institute for Advanced Studies. 

\section*{}

\begin{thebibliography}{31}
\bibitem{note} At zero electric field, there are three degenerate states corresponding to $N=1$ and $M_N=0, \pm1$. 
In the presence of an electric field relevant for this study, the degeneracy is lifted. As shown in Ref. \cite{felipe}, the $|N=0 \rangle \rightarrow |N=1, M_N = 0 \rangle$ transition is uncoupled from the $|N=0 \rangle \rightarrow |N=1, M_N = \pm 1 \rangle$ transitions. The $M_N = \pm 1$ states can, therefore, be ignored.
\bibitem{note1}
This analysis assumes that both $D$ and $J$ are isotropic. For molecules on an optical lattice, the magnitude and sign of $D$ and $J$ depend on the angle between the direction of the electric field and the molecular array. This may modify the properties of biexcitons in 2D and 3D systems.
\end{thebibliography}
\bibliographystyle{unsrt}

\end{document}